\documentclass[12pt]{iopart}

\usepackage{iopams}
\usepackage{graphicx}

\newcommand{\gr}{\gamma_{\mathrm{rad}}}
\newcommand{\gnr}{\gamma_{\mathrm{nonrad}}}
\newcommand{\gt}{\gamma_{\mathrm{tot}}}

\begin{document}

\title[Nanodiamond NV centers]{Suitability of nanodiamond NV centers for spontaneous emission control experiments}

\author{Abbas Mohtashami and A Femius Koenderink}

\address{Center for Nanophotonics, FOM Institute for Atomic and Molecular Physics (AMOLF), Science Park 104, 1098 XG Amsterdam, The Netherlands}
\ead{a.mohtashami@amolf.nl}
\begin{abstract}
NV centers in diamond are generally recognized as highly promising as indefinitely stable highly efficient single-photon sources. We report an experimental quantification of the brightness, radiative decay rate, nonradiative decay rate and quantum efficiency of single NV centers in diamond nanocrystals. Our experiments show that the commonly observed large spread in fluorescence decay rates of NV centers in nanodiamond is inconsistent with the common explanation of large nanophotonic mode-density variations in the ultra-small high-index crystals at near-unity quantum efficiency. We report that NV centers in 25\,nm nanocrystals are essentially insensitive to local density of states (LDOS) variations that we induce at a dielectric interface by using liquids to vary the refractive index, and propose that quantum efficiencies in such nanocrystals are widely distributed between 0\% and 20\%.   For single NV centers in larger 100\,nm nanocrystals, we show that decay rate changes can be reversibly induced by nanomechanically approaching a mirror to change the LDOS\@. Using this scanning mirror method, for the first time we report calibrated quantum efficiencies of NV centers, and show that different but nominally identical nanocrystals have widely distributed quantum efficiencies between 10\% and 90\%.  Our measurements imply that nanocrystals that are to be assembled into hybrid photonic structures for cavity QED should first be individually screened to assess fluorescence properties in detail.
\end{abstract}

\maketitle

\tableofcontents

\section{Introduction}

The large promise of quantum-optical technologies to enable secure communication and novel computation architectures sets  stringent targets for the quality of  {photon} sources and detectors. In order to meet these  demands many efforts are currently devoted to realizing bright sources of single photons~\cite{Santoribook,Vahalabook}. These developments at the same time require novel nanophotonic engineering designs around emitters to enhance light-matter interaction strength    as well as indefinitely stable two-level systems that neither bleach, blink, nor spectrally jump.  On the nanophotonic engineering side,  many different systems have been proposed to control whereto, how fast and with which  polarization an emitter emits provided that one manages to locate it exactly in the right location. These systems include whispering-gallery-mode cavities~\cite{Vahalabook}, micropillars and cylindrical wires~\cite{Lalanne,Gerard_JLT_1999,Wurzburg}, photonic crystal microcavities~\cite{Yoshie,Vuckovic,Noda},  photonic crystal waveguides near cut-off ~\cite{Hughes,LodahlWG},  Anderson localizing systems~\cite{LodahlAL},  as well as ultrabroadband plasmonic waveguides and antennas~\cite{ChangLukin,OultonZhang,Koenderink2009,Curto}.  Essentially, all these techniques  modify  the photonic environment of an emitter via the   Local Density of Optical States (LDOS)~\cite{Sprik1996,Novotnybook}.  The LDOS quantifies the light-matter interaction strength that appears in Fermi's Golden Rule for spontaneous emission.  Placing the emitter at a position where the LDOS is enhanced, first implies a much higher fluorescence decay rate. Second, if the LDOS enhancement is due to a select set of designed modes as in a cavity or waveguide,  the enhanced rate is accompanied by extraction of photons preferentially via these  enhanced modes. Thus, photonic engineering promises optimum brightness, dynamics, and collection efficiency. As regards the choice of emitter,  a wide variety of systems have been used. Unfortunately, all choices appear to carry large disadvantages when going beyond pilot studies: most dye molecules photobleach \cite{Lounis2005},  II-VI quantum dots~\cite{bawendi,vanmaekelbergh}  blink as well as bleach,  and many systems, such as III-V emitters, only have desirable properties when cooled to cryogenic temperatures ~\cite{Lalanne,Gerard_JLT_1999,Yoshie,Vuckovic,Wurzburg,LodahlWG,LodahlAL}. A promising candidate to provide an indefinitely stable source~\cite{Brouri_OL_2000,Kurtsiefer_PRL_2000,Beveratos_PRA_2001} that furthermore allows room-temperature spin-control~\cite{vanderSarnature,Robledo,Stanwix,PNASspinreadout,Tisler_ACSnano_2009}  is the nitrogen-vacancy (NV) color center in diamond.

When one chooses diamond NV centers as emitters for quantum optics in nanophotonic devices, one can either aim to manipulate emission by fabricating photonic structures directly in diamond~\cite{Faraon_PRL_2012,Santori_nanotech_2010,Babinec}, or one can assemble diamond nanocrystals with photonic structures of a different material ~\cite{NJP13_055017,Fu_NJP_2011,Faraon_NP_2011,Benson_Nature_2011,Schietinger_NL_2009,vanderSar_APL_2011,Kolesov_NPh_2009,Huck_PRL_2011}. Recently, pick-and-place strategies~\cite{Benson_Nature_2011,Schietinger_NL_2009,vanderSar_APL_2011} were reported in which a single nanoparticle  from a diluted powder of diamond nanocrystals dispersed on a substrate is selected and pushed to a desired location by a scanning probe, such as an atomic-force-microscopy tip, or a manipulator in a scanning electron microscope.  Reports of this technique span from the  coupling of nanodiamonds to  photonic-crystal cavities~\cite{Schietinger_NL_2009,vanderSar_APL_2011} to coupling to plasmonic antennas and nanowires~\cite{Kolesov_NPh_2009,Huck_PRL_2011}. As a variation of the pick-and-place strategy, a few  groups recently developed  so called `scanning-emitter'  near-field microscopes in which nanosources are not deposited irreversibly inside a nanostructure, but actually remain attached to a scanning- probe~\cite{Michaelis2000,Cuche2009,Cuche2010,FrimmerPRL}.   The advantage of such a scanning probe approach is that one can first construct a full map of the LDOS using near-field lifetime imaging to determine where one should ultimately place the nanosource~\cite{FrimmerPRL}.  Microscopy with a light-source as a near-field tip is interesting as a microscopy technique, but only viable if the nanosource is indefinitely stable and does not blink, for which nanodiamonds appear the sole candidate~\cite{Cuche2009,Cuche2010}.  Moreover nanodiamond scanning probes offer the possibility to create nanometer sized local magnetometry probes read out optically by NV spectroscopy~\cite{Degen2008,Taylor2008,Balasubramaniam2012,Maletinsky2012}.

In view of the scanning probe microscopy  and assembly efforts that seek to combine nanodiamonds and photonic structures,  one  question stands out as of key importance:  given an ensemble of individual nanodiamonds, how does one recognize the  ideal nanocrystal?  This question is especially relevant given that both fabrication of the photonic structure \emph{and} the intended  scanning-probe procedure is highly laborious~\cite{Benson_Nature_2011}.  Naively, one might think that  all nanocrystals of subwavelength size that contain a single NV center will be equally suited,  since an NV center is a  defect of uniquely defined composition and geometry in the diamond crystalline matrix.  However,  many workers on diamond nanocrystals have established that NV centers in diamond nanocrystals actually show a  distribution of photophysical properties, such as brightness, stability, and decay rates~\cite{Tisler_ACSnano_2009,Schietinger_NL_2009,NL9_3555,Inam_NJP_2011,IOPNAnoTech_23_175702,Ruijgrok_OE_2010}. This distribution is usually ascribed to the fact that even though all NV centers are expected to have unit quantum efficiency and the same oscillator strength, these identical unit-quantum-efficiency sources are each differently placed  inside their nanoscopic high-index diamond grains~\cite{Brouri_OL_2000,Kurtsiefer_PRL_2000,Beveratos_PRA_2001,Inam_NJP_2011,Ruijgrok_OE_2010}. This is anticipated to cause different decay rates, due to the fact that the LDOS even in an isolated nanoscopic object  varies as a function of position and dipole orientation~\cite{Schniepp}.

In this paper we address the question how to recognize the  ideal nanocrystal from an ensemble of nanocrystals on basis of the requirement that the ideal nanocrystal must contain a single NV$^-$ center that fluorescences with a high quantum yield, so that it can be useful as a reporting probe of LDOS in nanophotonic systems.  We report that the commonly made assumption that NV centers are unit-efficiency dipoles randomly distributed in high index nanoscopic objects is not reconcilable with measured brightness and decay rate histograms data for nanocrystals in the frequently used size ranges around 20--50\,nm and 50--150\,nm.     On basis of the wide distribution of brightness and rate that can not be explained by LDOS variations between nanoparticles alone we conclude that both the radiative and nonradiative rates are broadly distributed. To quantify this distribution, we have performed experiments in which we measure emission rate changes of individual nanodiamonds as we controllably vary the LDOS of their environment.  For small nanocrystals we find a decay rate distribution indicative of low-quantum-efficiency emitters. Furthermore, consistent with this low quantum efficiency, we find no evidence that small single NV centers are responsive LDOS probes when applying moderate LDOS variations.  For the larger nanocrystals (100 nm diameter), we   for the first time manage to induce reversible changes of up to 25\% in the total decay rate of single NV centers using a calibrated LDOS change induced by a nanomechanically moved mirror. On basis of these measurements we argue that the apparent quantum efficiency of nanodiamonds  in the size range 50--150\,nm  ranges widely from about 10\% to 90\%. This wide range of quantum efficiencies implies that prior to constructing a photonic structure, it is necessary to screen the nanocrystals using a calibrated quantum- efficiency measurement on the single NV center level, as shown in this paper. This paper is organized in the following manner. First we report on our experimental methods in section~\ref{sec:setup}.  Next, we discuss the distributions of brightnesses and rates for large nanocrystals in section~\ref{subsec:distriblarge} and for small nanocrystals in section~\ref{subsec:distribsmall}.  Finally we report on efforts to tune rates of NV centers in small diamonds by controlled LDOS changes. The two methods used are liquid tuning of LDOS in section~\ref{sec:liquidldos}, and nanomechanically approaching a mirror as reported in  section~\ref{sec:mirrorldos}.

\section{Experiment and methods}\label{sec:setup}

\subsection{Sample preparation}
\label{subsec:smaple_prep}
Since single NV centers in nanodiamonds are  comparatively dim emitters that require high excitation powers, it is essential to avoid background fluorescence when performing fluorescence microscopy. Therefore, we use intrinsically low-fluorescent quartz coverslips (Esco Products) as sample substrates that were cleaned by $15\,\mathrm{min}$ sonication in water followed by a $15\,\mathrm{min}$ bath in base Piranha (NH$_3$(aq.,$~30\,\%$)\,:\,H$_2$O$_2$(aq.,$~30\,\%$)\,:\,H$_2$O mixed at ratio 1:1:5,  at  $75\,^\circ$C). In our work it is essential that we can unambiguously pinpoint the position of nanodiamonds containing NV centers in the course of the measurement, so that we can revisit the same color center after, for instance, material-deposition steps. To this end we define an array of markers on top of the coverslips using electron-beam lithography. To remove any organic residues after the lift-off process, we treated the coverslips with a mild $\mathrm{O}_2$-plasma descum (Oxford Instruments Plasmalab 80+, using a 5\,mTorr, 25\,sccm $\mathrm{O}_2$ plasma).  As sources for NV centers, we use solutions of monocrystalline synthetic nanodiamonds (Microdiamant MSY) with a narrow size tolerance obtained from Microdiamant AG, Lengwil, Switzerland. We note that these nanoparticles have been used recently by a large number of groups in experiments that rely on fluorescence~\cite{Stanwix,PNASspinreadout,Schietinger_NL_2009,NL9_3555,NJP13_055017,IOPNAnoTech_23_175702,Huck_PRL_2011,Kolesov_NPh_2009,Inam_NJP_2011}.
We use these nanodiamonds exactly in the manner as described by Schietinger \etal, i.e.\ without further washing or purification steps \cite{Schietinger_NL_2009,SchietingerPrivcomm}. We prepared samples from diamond slurries with two different size distributions by spin-coating aqueous solutions of nanodiamonds that are diluted to a concentration of 1\% of the as-received stock solutions on the cleaned and patterned coverslips. The first type of sample, from here on referred to as `25\,nm diamond sample'  was made from Microdiamant MSY 0--0.05, which nominally has sizes from 0 to 50\,nm. These diamonds have a median diameter of 26\,nm, with fewer than 1\% of the diamonds above 50\,nm in size, according to particle sizing performed on this batch of diamonds by the manufacturer. The second type of sample (`100\,nm diamond sample') was made from Microdiamant MSY 0--0.2, which nominally has crystal sizes ranging from 0 to 200\,nm. This ensemble has crystals  with median diameter 108\,nm, and fewer than 1\% of particles above 175\,nm in size, again according to the size-distribution histogram supplied by the manufacturer. Scanning electron microscope (SEM) inspection indicated a size distribution of nanodiamonds consistent with these specifications. The average density of spin-coated nanodiamonds on the coverslips as checked through the SEM images was about 1 to 8\,$\mu \mathrm{m}^{-2}$ depending on the sample, assuring that only a few nanodiamonds are illuminated at the diffraction-limited focus of the objective. Only a small fraction of the nanodiamonds are fluorescent, and an even smaller fraction fluoresces due to an NV center.  We identify a diamond as containing at least one NV center, if we can clearly identify the characteristic zero phonon line (ZPL)~\cite{Brouri_OL_2000,Kurtsiefer_PRL_2000,Beveratos_PRA_2001,Plakothnik} in its emission spectrum according to the criterion we specify in  section \ref{Single_NV_charac}.  With this criterion, for 100\,nm nanodiamonds at an average density of 1\,$\mu \mathrm{m}^{-2}$, we found on average one NV center in an area equal to $50\times 50$\,$\mu \mathrm{m}^{2}$, while for 25\,nm nanodiamonds at an average density of 5\,$\mu \mathrm{m}^{-2}$ we found on average 2 NV centers in a $100\times 100\mu$\,$\mathrm{m}^{2}$ area. These numbers translate to identifying fewer than 0.05\% of the 100\,nm nanodiamonds, and  fewer  than 0.001\% of the 25\,nm diamonds as containing  a (single) NV center beyond doubt.  Schietinger \etal~\cite{Schietinger_NL_2008} have reported a  higher density of NV centers of about 1\% for MSY 0--0.05 nanodiamonds.  A difference in reported  NV center densities could be either due to a different degree of  strictness in labeling a fluorescent emitter as an NV center beyond doubt (see section~\ref{Single_NV_charac} for our criteria), or alternatively to batch-to-batch variations in nanodiamond slurries.

\subsection{Experimental setup}
\label{subsec:Exp_setup}

\begin{figure}[t]
\begin{center}
 \includegraphics[width=80mm]{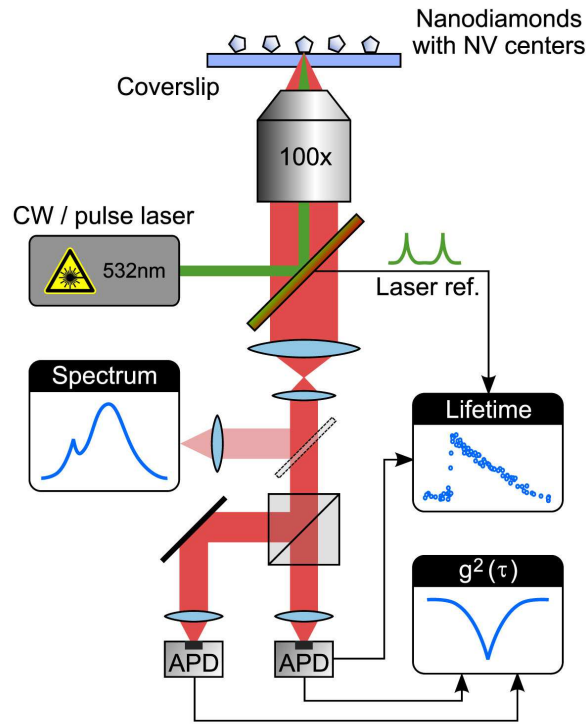}
 \caption{Schematic of the experimental setup. Spin-coated nanodiamonds on quartz coverslip are illuminated by 532\,nm laser through a $100\times$ dry objective. The emitted fluorescence can be guided to spectrometer or to APDs to measure the lifetime and second order correlation function $g^{(2)}$ of the emitted photons. \label{Fig:Setup}
}
\end{center}
\end{figure}

The optical setup, sketched in figure~\ref{Fig:Setup}, consists of an inverted confocal fluorescence microscope equipped with a sample scanning piezo stage. NV centers are optically excited at 532\,nm using either a frequency-doubled Nd:YAG pulsed laser (Time-Bandwidth, 10\,MHz repetition rate) or a continuous-wave (CW) diode laser (CNI). The laser beam is focused through the coverslip to a diffraction-limited spot on top of the sample, using a $100\times$ dry objective with a numerical aperture of 0.9 (Nikon CFI Plan Fluor). The same objective collects the luminescence and guides it through a long-pass filter (580\,nm cut-off) before it is imaged on a CCD camera (Nikon DS-Qi1Mc) or confocally detected on one or two  avalanche photodiodes (APD) with single-photon sensitivity (both APDs: id Quantique id100-20ULN). The APDs in combination with a sub-nanosecond-resolution 16 channel correlator (Becker \& Hickl, DPC-230) allow for single-photon counting. The correlator can perform time-correlated single-photon-counting (TCSPC) lifetime measurements by correlation of detection events with laser pulse arrival times, or photon-photon correlations using multiple APDs. We use a second APD in a Hanbury-Brown and Twiss (HBT) configuration to measure photon correlation statistics (antibunching) at CW excitation. Single-nanocrystal spectra are collected using an imaging spectrometer (SpectraPro 2300i) equipped with a thermoelectrically cooled back-illuminated Si CCD array detector (Princeton Instruments PIXIS:100B).

\subsection{Measurement procedure to characterize single nanodiamonds}
\label{Single_NV_charac}

\begin{figure}[t]
 \includegraphics[width=\textwidth]{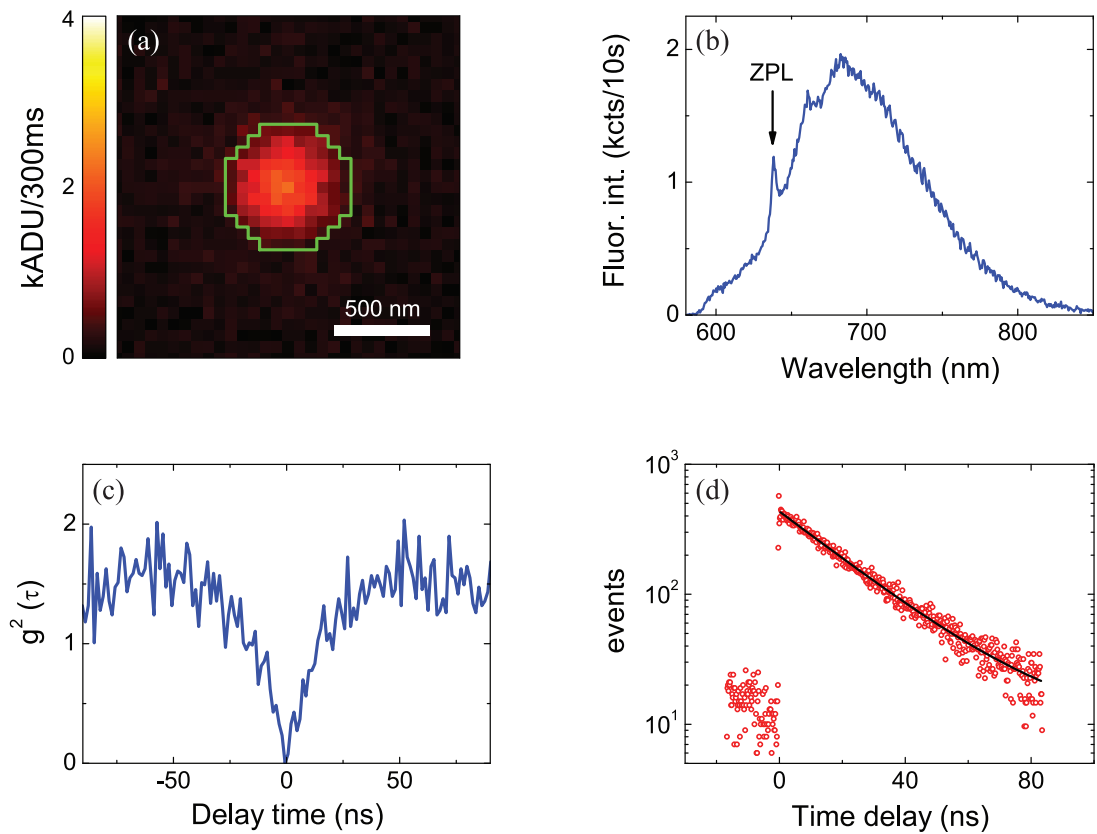}
 \caption{(a) CCD image of the fluorescence from a potential NV center positioned at the focus of the laser. Green line indicates an integration area over which the fluorescence intensity of the emitter is evaluated. (b) Typical spectrum of an NV center expressing the characteristic zero phonon line at 638\,nm. (c) Second-order correlation function $g^{(2)}$ of an NV center obtained by CW excitation. The zero dip in $g^{(2)}$ curve confirms the single nature of the center.   (d) Background-corrected decay traces of photons emitted from an NV center (red circles, total acquisition time 20 s) with a single-exponential fit (black curve) to extract the lifetime. \label{Fig:Fig_1}
}
\end{figure}

In order to find nanodiamonds containing NV centers, the sample is first pumped with wide-field laser illumination and the fluorescence is imaged on the CCD camera.  Under these conditions, any  potential NV center will appear as a diffraction-limited fluorescent spot. Subsequently, we switch to diffraction-limited illumination and  position  the potential candidate at the laser focus using a piezostage (figure~\ref{Fig:Fig_1}(a)). The fluorescence intensity of the emitter is quantified by integration over a fixed CCD area which is indicated by the green line in figure~\ref{Fig:Fig_1}(a). At this stage, we spectrally resolve the fluorescence from the isolated emitter with the spectrometer to specifically find the characteristic zero phonon line of NV centers at a wavelength of about 638\,nm (figure~\ref{Fig:Fig_1}(b))~\cite{Brouri_OL_2000,Beveratos_PRA_2001,Kurtsiefer_PRL_2000}.  On basis of the ZPL, we either discard the emitter (absence of ZPL), or identify the nanodiamond as having one or more NV centers (ZPL well above the detector and photon-counting noise). For each identified NV center, we record a second-order correlation function $g^{(2)}(\tau)$  to verify that the emission originates from a single center. To this end, we pump the center with the CW laser at a power of about 1\,mW to achieve count rates on the order of about $5\times 10^4$ counts per second on each APD and integrate for about 1000\,s to obtain reasonable correlation statistics of around 30 coincidences per 658\,ps bin. Figure~\ref{Fig:Fig_1}(c) shows a typical background-subtracted $g^{(2)}(\tau)$ curve, with a clearly resolved minimum at zero delay $\tau=0$ that  is well below 0.5, indicating that emission is from a single NV center. Here, the background  is subtracted via the procedure reported in \cite{Brouri_OL_2000}, based on a measurement directly next to the NV center where we expect identical  fluorescence background from the substrate. The background measured  only from the substrate shows a Poissonian emission statistics. The  bunching in the $g^{(2)}(\tau)$ in figure~\ref{Fig:Fig_1}(c)  at longer delay times (visible at $\tau \approx \pm$~50\,ns) is well known to be due to the presence of a shelving state \cite{Kurtsiefer_PRL_2000}. Once we have identified a nanodiamond as containing a single NV center based on its spectrum and antibunching signature, we measure its fluorescence lifetime by switching to pulsed laser excitation. The lifetime of each center is extracted by fitting a single-exponential decay incorporating a small constant background to the decay histogram of emitted photons as shown in figure~\ref{Fig:Fig_1}(d). We note that the substrate itself has a weak background fluorescence that shows a time dependence. We correct for this artifact by collecting decay traces from the sample pumped directly next to the NV center. We parameterize the background by a tri-exponential fit (with a dominant sub-nanosecond lifetime) which we subtract from the  NV center decay histogram prior to fitting. The residual background is around 10 counts per bin  for a total acquisition time of about 20\,s.

\section{Statistics on fluorescence parameters  of  large nanodiamonds}

In this section we report on statistical distributions of the various fluorescence characteristics, i.e.\ brightness, $g^{(2)}$ and lifetime, that we collected on both nanodiamond size ranges. We first discuss the `100\,nm diamond' sample, containing  0--200\,nm size nanodiamonds with 108\,nm median diameter, and subsequently we discuss the `25\,nm diamond' sample with  0--50\,nm size nanodiamonds and 26\,nm median diameter. For each sample, our statistics is based on identifying 30 to 40 NV centers as described in section \ref{Single_NV_charac}.

\subsection{Distribution of brightness and relation to $g^{(2)}(0)$}
\label{subsec:distriblarge}
\begin{figure}[t]
 \includegraphics[width=150mm]{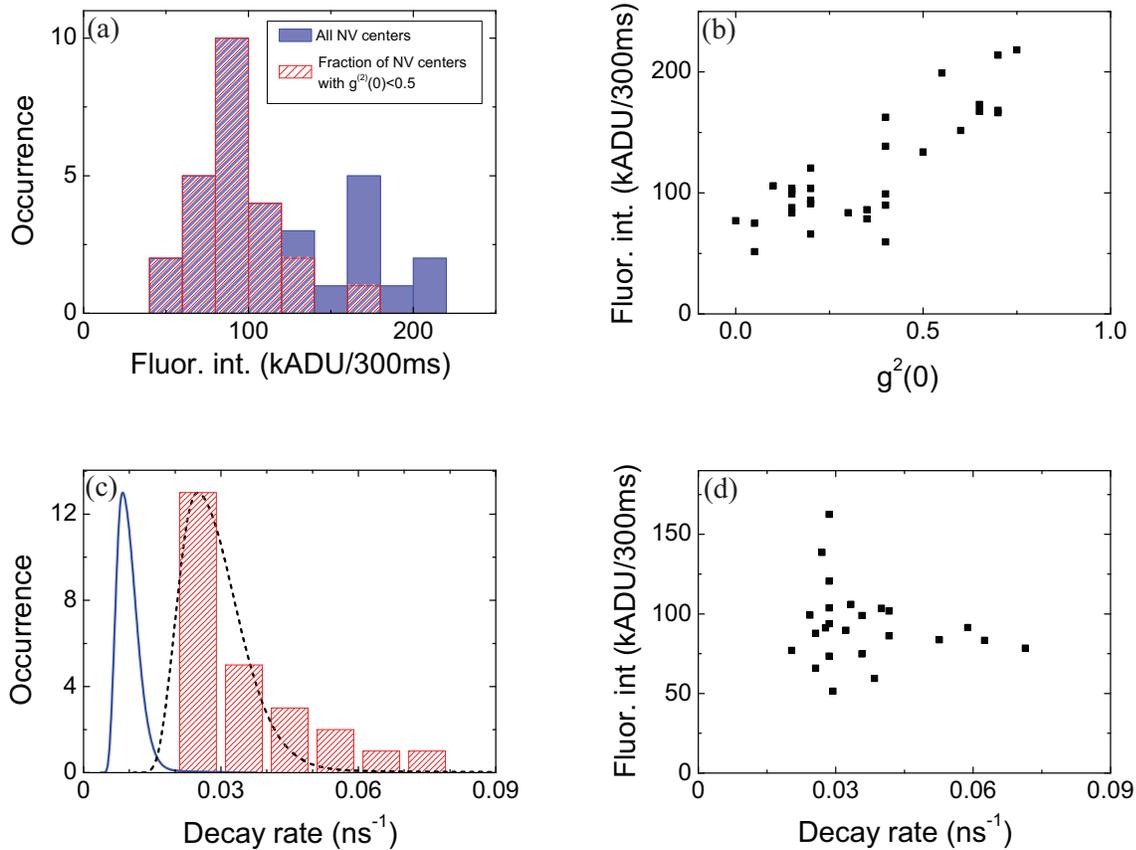}
 \caption{Statistics for 100\,nm nanodiamonds. (a) Histogram of fluorescence intensity of NV centers. Solid bars indicate the fluorescence distribution of all identified NV centers whereas pattered bars are related to the fraction of single NV centers identified by $\textrm{g}^{2}(0)<0.5$ in (b)\@. (b) Fluorescence intensity as a function of $\textrm{g}^{2}(0)$\@. (c)  {Bars:} histogram of decay rates of emitters with $\textrm{g}^{2}(0)<0.5$. Solid line: calculated distribution of $\textrm{LDOS}/\textrm{LDOS}_{vac}$  according to nanodiamond size distribution, normalized to the rate of NV centers in bulk diamond. Dotted line: free scaling of the calculated LDOS distribution (solid line)   to match the experimental histogram (bars)\@. (d) Fluorescence intensity as a function of decay rate for emitters with $\textrm{g}^{2}(0)<0.5$  (decay rate error bars are within symbol size). Error bars are comparable to plot symbol size.\label{Fig:Fig_2}}
\end{figure}

As a first step in the characterization process, we have measured the brightness of nanodiamonds that show a
clear ZPL line. Figure~ \ref{Fig:Fig_2}(a) shows a histogram of measured brightnesses as quantified by intensity on the CCD at fixed illumination intensity (solid bars). The intensity on the CCD is obtained by summing all pixels within the diffraction-limited image of each diamond (green line in figure~\ref{Fig:Fig_1}(a)). We observe a wide distribution of intensities, spanning from $50\cdot 10^3$ to $210\cdot 10^3$ ADU's (Analog Digital Units) on the CCD per 300\,ms exposure time. At an estimated photon-to-ADU conversion factor of 6 for our CCD camera, these brightnesses correspond to $1\cdot 10^6$ to $4.2\cdot 10^6$  collected photons per second, at a pump power around $1 \mathrm{mW}$ supplied by the CW laser.   First, these numbers show, at least assuming that NV centers are reasonably efficient emitters, the absorption cross sections of nanodiamond NV centers at $532$\,nm pump wavelength are an order of magnitude  below  those of dyes and II-VI quantum dot nanocrystals.  Second, the wide  distribution of brightnesses of over a factor 3 to 4  from nominally identical emitters is highly surprising. As described in section \ref{Single_NV_charac}, some nanodiamonds may have multiple NV centers, especially in the case of big nanoparticles. To exclude that brightness variations are due to multiple NV centers, we screen the nanodiamonds on basis of $g^{(2)}$ minimum values. Figure~\ref{Fig:Fig_2}(b) shows a  correlation plot, plotting the fluorescence intensities of all identified NV centers on 100\,nm nanodiamond sample versus the recorded value of $g^{(2)}$ at zero delay ($g^{(2)}(\tau=0)$). We find that not just the brightness, but also the minimum in $g^{(2)}$ is distributed with minima in $g^{(2)}$ ranging from 0 to 0.75. The minimum in $g^{(2)}(0)$ is only weakly correlated with collected intensity, especially via an apparent stepwise increase in the fluorescence intensity as $g^{(2)}(0)$ reaches above $\approx 0.5$. Such a stepwise increase would be expected since the minimum in $g^{(2)}$   scales with the number of emitters $n$ as $1-1/n$. For $g^{(2)}(0)\gtrsim 0.5$ more than a single fluorescent center is involved in the emission, resulting in higher fluorescence intensities on average.

In the remainder we concentrate our analyses on  single NV centers, i.e.\  those nanocrystals for which $g^{(2)}(0)<0.5$.  In figure~ \ref{Fig:Fig_2}(a) , we overplot a histogram of fluorescence intensities of the subset of centers with $g^{(2)}(0)<0.5$  as patterned bars. The histogram shows a wide fluorescence distribution, with a factor of three difference between brightest and dimmest NV centers, and with a relative distribution of about 30\% around the most frequent value. Figure~\ref{Fig:Fig_2}(b)  shows that  for the single NV centers the wide distribution in  fluorescence intensities   notably does not show any correlation to $g^{(2)}(0)$ values.  Previously, Beveratos \etal  \cite{Beveratos_PRA_2001} attributed the distribution of $g^{(2)}(0)$ values  to inseparable background emission from the nanodiamond crystal host. The lack of correlation between brightness and $g^{(2)}(0)$ suggests that this assertion is not valid, i.e. that differences in the NV emission itself causes the brightness distribution.  It should be noted that  brightness differences between NV centers  could occur due to the fact that the linearly polarized pump field projects differently on the NV center absorption dipole moment. According to Ref.~\cite{Alegre}, an NV center has two dipole moments of equal size in the plane perpendicular to the NV axis. For a random distribution of NV orientations $\mathbf{p}$, the dipole projection $|\mathbf{p}\cdot\mathbf{E}|^2$ on a linearly polarized pump field $\mathbf{E}$  would lead to a brightness distribution that is strongly asymmetric, skewed strongly  to high brightness. In contradistinction, the brightness histrogram we find for NV centers is not skewed to high brightness. To conclude, the single NV centers display a variation in brightness over a factor 3, which can not be attributed to either background emission of the crystal host or variations in absorption cross-section through dipole orientation.

\subsection{Correlation of emission rates and brightness}
On basis of the large brightness  distribution that can neither be explained by dipole orientation nor by background fluorescence,  we hypothesize that there is a substantial variation in  quantum efficiency (QE) of the NV centers. To examine whether quantum efficiency effects may be at play, we have measured fluorescence decay rates for each identified single NV center. We plot  a histogram of the measured decay rates in figure~\ref{Fig:Fig_2}(c). We observe a very wide distribution, with the slowest emitters decaying almost 4 times more slowly than the fastest ones.  The most frequently occurring decay rate is around $\gt=0.03$\,ns$^{-1}$ (corresponding to  about 33\,ns).   The time constant of 33\,ns is appreciably slower than the fluorescence lifetime of NV centers  in bulk diamond, for which the accepted literature value is 11.6~ns~\cite{Beveratos_PRA_2001}.  This much slower decay as well as the occurrence of a distribution of rates is in agreement with previously reported values \cite{Inam_NJP_2011,Ruijgrok_OE_2010}. If the hypothesis that  quantum efficiency variations are  responsible for the large variability in brightness in figure~\ref{Fig:Fig_2}(a) is valid, one might expect a correlation between the brightness and decay rate of the emitters.
We plot the measured brightness as a function of decay rate for each NV center in figure~\ref{Fig:Fig_2}(d) which, however, displays no clear correlation. Importantly, we note that the measured quantity here is the total decay rate $\gt=\gr+\gnr$ which reflects both variations of radiative decay rate $\gr$ and nonradiative decay rate $\gnr$, whereas the quantum efficiency is given by $\gr/\gt$.
Based on the collection efficiency of our microscope objective ($\sim 10$ \%), the count rates in Figure~\ref{Fig:Fig_2} correspond to an intermediate (most data) to strong (data point at and above 150 kADU/300 ms))excitation regime, where the photon emission rate is approximately 0.2 to 0.9 times the total decay rate $\gt$. Both if one assumes to be in saturation, and below saturation, one expects  the fastest decay  to imply highest brightness  if   one assumes $\gnr$ to be approximately constant but $\gr$ to be distributed. This conclusion is not strongly supported by the data.  If conversely we assume $\gr$ to be approximately constant across NV centers while $\gnr$ is distributed, one would expect highest  brightness to correlate with the slowest decay rate (lowest $\gnr$), a hypothesis also not strongly supported by the data.    Specifically, we find that the subset of crystals around  the most frequent brightness show the full spread of decay constants, while conversely also the subset of crystals that have decay constant around the most frequent rate, contain the full range of brightnesses. We  conclude that the common assumption that  NV centers in such large nanocrystals as we study are near-unity quantum efficiency sources is not supported by the data. Instead  \emph{both} the radiative and the nonradiative decay rates could be widely distributed.

\subsection{Common LDOS argument for rate variation in nanodiamond}
Many workers had already noticed that nanodiamond decay rates are widely distributed \cite{Inam_NJP_2011,Ruijgrok_OE_2010,Tisler_ACSnano_2009}. The variation is commonly attributed to variations solely in the radiative rate $\gr$ due to a local-density-of-photonic-states effect, assuming zero nonradiative decay (i.e.\ unit quantum efficiency).
The hypothesis, explained in detail by Inam \etal \cite{Inam_NJP_2011}, is that variations are in large part due to the fact that the LDOS experienced by an NV center is influenced by the nanoscale geometry  of its environment, i.e.\ the fact that the source is situated inside, and close to the surface of, a very high index nano-object that is embedded in a low index environment. A distribution in LDOS can arise from the size distribution of nanodiamonds and from the fact that different NV centers have different  dipole-moment orientation and positions within the crystals~\cite{Schniepp,Inam_NJP_2011}. Here we assess if this LDOS hypothesis is quantitatively reasonable.
 Considering that the radiative decay rate $\gr(\rho)$ is proportional to LDOS, we  evaluate the $\gr(\rho)$ distribution by calculating the LDOS distribution assuming spherical nanoparticles, for which the LDOS is analytically known~\cite{Taibook,MertensPRB2007}. We modeled nanodiamond NV centers as point dipoles with randomly oriented dipole moments  homogeneously distributed in position in a set of dielectric spheres with  dielectric constant equal to the bulk diamond value ($\varepsilon_{\mathrm{sphere}}=5.85$). We furthermore take into account the distribution of particle size  as specified  according to the nanodiamond size-distribution histogram provided by the manufacturer. We histogram the occurrence of  LDOS values to find its probability distribution over all sphere sizes, dipole positions and dipole configurations.
The resulting LDOS distribution can be converted into a distribution directly comparable to experimentally measured decay rates by scaling the  LDOS to the previously reported  decay  rate of  (11.6~\,ns)$^{-1} \approx 0.086$\,ns$^{-1}$ of NV centers in bulk diamond~\cite{Beveratos_PRA_2001}. We note that this entire procedure involves no adjustable parameter or any fit to data. The final result is plotted in figure~\ref{Fig:Fig_2}(c) as the blue solid line.  The calculated histogram correctly predicts that emission is significantly decelerated compared to decay in bulk diamond, consistent with the fact that decay in small dielectric spheres is decelerated both compared to vacuum, and bulk dielectric. However,  we observe that the calculated histogram peak falls at much lower decay rate than the experimentally measured histogram peak, with a discrepancy amounting to a factor of 3. A similar discrepancy between the calculated and experimentally measured decay rates of nanodiamond NV centers was recently reported for samples with a much wider size distribution, centered at much larger median size, measured by Inam \etal \cite{Inam_NJP_2011}.  We note that this discrepancy cannot be attributed to the fact that we have taken particle shape to be simply spherical, and that we have neglected the presence of a substrate. These effects cause only small changes in the expected decay rate histogram, as verified in FDTD simulations by Inam \etal \cite{Inam_NJP_2011}.  One might further argue that the occurrence of a degenerate in-plane dipole moment~\cite{Alegre} could skew the histogram of expected decay rates towards higher values, if one assumes the dipole moment is free to diffuse prior to de-excitation. However, if we just  histogram the fastest rate instead of the average rate at each possible NV center position, the resulting histogram  also does not lead to a consistent explanation (not shown). We conclude that LDOS variations in nanodiamond under the hypothesis of unit quantum efficiency and a bulk decay rate of (11.6 ns)$^{-1}$  do not explain the variation in measured decay rates.

\subsection{LDOS argument beyond unit quantum efficiency}
Inam \etal proposed that  although calculated and experimental absolute decay values are inconsistent on basis of LDOS theory and the bulk rate in diamond,  the calculated and experimental results can be scaled onto each other.  Indeed, if we scale the reference rate that sets the rate axis for  the calculated histogram peak not by 11.6\,ns (rate constant in bulk diamond~\cite{Beveratos_PRA_2001}), but by a factor of 3 shorter, calculated and measured histogram coincide reasonably (figure~\ref{Fig:Fig_2}(c) black dotted line).
Such a scaling would imply as hypothesis a  unit quantum efficiency,  together with an as yet hidden explanation that introduces a multiplicative correction factor in the calculated rate distribution.

A second hypothesis could be that no adjustment should be made of the bulk rate that enters the comparison, but that the quantum efficiency of NV centers, while unity in bulk, is not unity in nanocrystals. Indeed,  an additive offset to the calculated histogram is introduced by  nonradiative decay channels that do not occur for bulk diamond, but could occur for nanocrystals due to defects and the presence of large surface area that could contain quenching sites. So far, significant nonradiative decay was evidenced only for very small nanocrystals (5\,nm~\cite{Plakothnik}).  A distribution of nonradiative rates around $\gnr= 0.15$\,ns$^{-1}$ would shift  the calculated histogram to the measured rates. The magnitude of the required $\gnr$ implies  that quantum efficiencies should be  around  30\% to 50\% for the slowest nanocrystals in the measured ensemble (assuming unit efficiency in bulk).

A third hypothesis could be that the assumption that the bulk rate is entirely radiative to start with, is incorrect, i.e.\ that the quantum efficiency of NV centers in bulk is significantly below unity contrary to common assumption. Assuming a bulk quantum efficiency of 70\% instead of 100\% would overlap the calculated histogram peak with the measured most frequent rate. However, this explanation would severely underestimate the width of the decay rate distribution unless a distribution of $\gnr$ is at play.

We conclude  that LDOS variations in nanodiamond alone do not explain the variation in measured decay rates, and that a distribution of radiative and nonradiative decay constants must be at play for NV centers in nanodiamond. Furthermore we conclude that an actual experimental calibration of quantum efficiency of individual NV centers is highly desired, which we will return to in sections~\ref{sec:liquidldos} and~\ref{sec:mirrorldos}.

\section{Statistics on small nanodiamonds}
\label{subsec:distribsmall}

\begin{figure}[t]
\includegraphics[width=150mm]{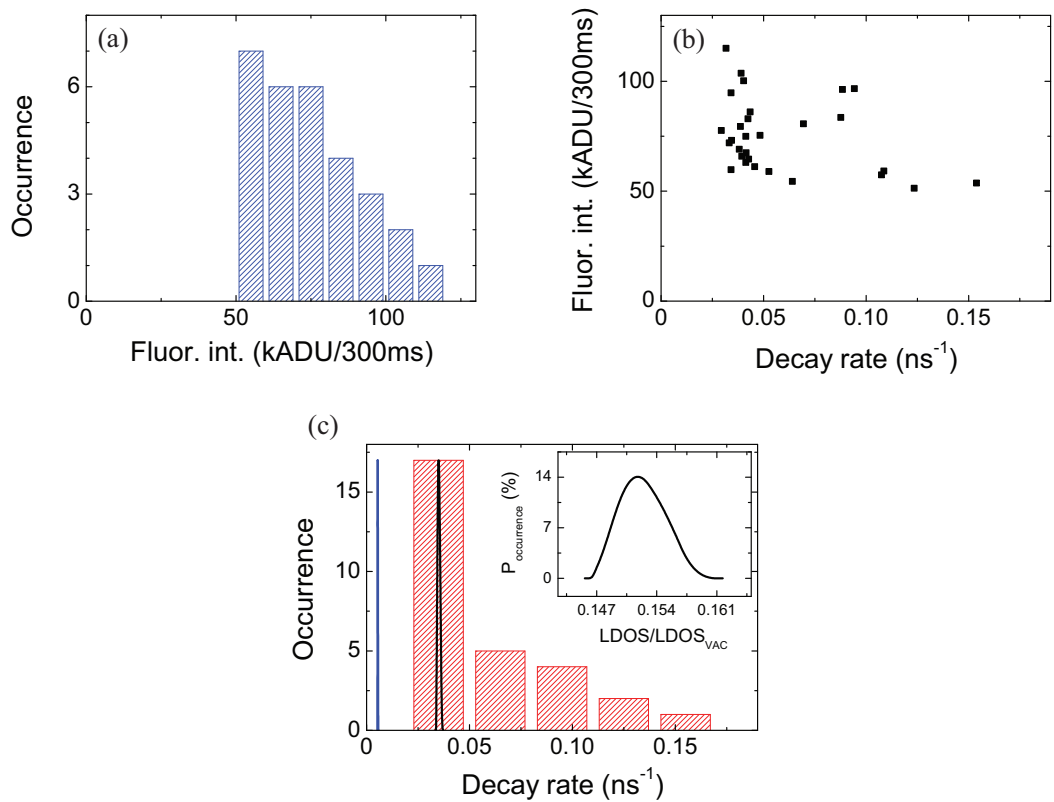}
 \caption{Statistics for 25\,nm ND. (a) Histogram of fluorescence intensity. (b) Fluorescence intensity as a function of decay rate. (c) Bars: histogram of the measured decay rates of NV centers. Solid blue line: calculated distribution of $\textrm{LDOS}/\textrm{LDOS}_{vac}$ (shown in the inset) according to nanodiamonds size distribution, normalized to the rate of NV centers in bulk diamond. Solid black line: free scaling of the calculated LDOS distribution (blue line) so to match the experimental histogram (bars)\@.  \label{Fig:Fig_3}}
\end{figure}

\subsection{Brightness and $g^{(2)}(0)$}
For experiments in which nanodiamonds are intended as probes of, or sources to be embedded in, nanophotonic environments, a size  smaller than that of the 100\,nm nanodiamonds  would be advantageous, as smaller size implies higher spatial resolution and a lower perturbative effect on the modes of the nanophotonic system~\cite{Koenderink_PRL_2005,Sar_JOSAB_2012}. Therefore, we repeated the brightness and decay rate statistical measurements for the  25\,nm sample, i.e.\ the batch of crystals with median diameter 26\,nm. For 25\,nm nanodiamonds, background-subtracted $g^{(2)}$ measurements show a zero dip for all identified NV centers (for a typical 25\,nm nanodiamond $g^{(2)}$ see e.g. figure~\ref{Fig:Fig_1}(c)). The fact that we found no nanodiamonds with multiple NV centers for this sample is commensurate with the smaller average crystal size. For these single NV centers, figure~\ref{Fig:Fig_3}(a) shows a histogram of the fluorescence intensities measured on the CCD using CW illumination at 1~mW/$\mu$m$^2$. As in the case of the  100\,nm sample, the histogram exhibits a wide fluorescence intensity distribution, in this case spanning a factor of about two.  The most frequent brightness is approximately a factor two below that of the 100\,nm sample.  The fact that the histogram is skewed towards lower intensities suggests that the intensity distribution is not due to the projection of the dipole moment orientation on the pump-field polarization, as this would cause a skewing in the opposite direction.

\subsection{Distribution of rates}
As in the case of the 100\,nm nanocrystals, we have also measured the decay rate for all 25\,nm nanocrystals that we identified.  The decay rate distribution, plotted as a histogram in figure~\ref{Fig:Fig_3}(c),  again  shows a very wide distribution, with a factor of 4 contrast in decay rate.  We find higher decay rates on average compared to 100\,nm nanodiamonds, with a most frequent decay rate around  $\gt=0.035$\,ns$^{-1}$ (corresponding to about 28\,ns).  The most frequent decay rate is approximately   25\%  faster than for the 100\,nm nanodiamonds.  Exactly as in the case of 100\,nm nanodiamonds, the wide distribution of  fluorescence brightness and decay rates does not imply a correlation between the two.   As confirmation, in figure~\ref{Fig:Fig_3}(b) we plot  the fluorescence intensities and decay rates of 25\,nm diamond NV centers,  where  we find  no clear correlation. We hence conclude that as for the larger diamonds,  a distribution in absorption cross-sections, radiative and nonradiative rates is likely at play.

\subsection{Comparison to LDOS argument for distribution of rates}
To assess whether the common hypothesis that the distribution in rate is due to a distribution in LDOS is valid for small nanodiamonds, we calculated the LDOS distribution (plotted in the inset of figure~\ref{Fig:Fig_3}(c))  also for the 25\,nm sample taking into account the size distribution histogram measured by the nanodiamond supplier. The rate distribution expected from the calculated LDOS scaled with the bulk rate (blue solid line in figure~\ref{Fig:Fig_3}(c)) is first, considerably narrower than the experimental decay rate distribution, and second, at considerably reduced rate compared to the measured decay rates.  The magnitude of the discrepancy in decay rate is approximately a factor 6, i.e.\ twice larger than for the 100\,nm nanodiamonds. A scaling of the calculated distribution by a multiplicative factor  (here a factor of about 6) as proposed by Inam \etal \cite{Inam_NJP_2011}  does not lead to a good correspondence as in the 100\,nm case, as the relative width of the measured histogram far exceeds that of the scaled calculation (black line in figure~\ref{Fig:Fig_3}(c)). Taking our data on both 100\,nm and 25\,nm nanodiamonds together, we hence do not find support for the hypothesis by Inam \etal that a hitherto hidden effect multiplies the radiative rate of NV centers in nanocrystals compared to bulk diamond. A more likely explanation that does not involve a scaling of $\gr$ due to an unknown origin is that NV centers are  subject to a distribution of nonradiative rates on top of the LDOS-induced radiative rate distribution.  The measured distribution of 25\,nm nanodiamond decay rates points at a wide distribution of nonradiative rates .  According to this hypothesis, those crystals with decay around the most-frequent decay rate of   $0.035$\,ns$^{-1}$ must have a quantum efficiency below 15-20\%, twice lower than for the 100\,nm nanocrystals.  The overall   lower quantum efficiency estimate is commensurate with the reduction in average brightness,  and is also consistent with the fact that a larger sensitivity to nonradiative decay channels is potentially  associated with the increased nanocrystal surface-to-volume ratio.

 \section{ LDOS tuning on single NV centers}
\label{sec:ldos}
Our data indicate  that NV centers even in  nanocrystals as large as 100 nm across do not have unit quantum efficiency,  and that there is a distribution of nonradiative and radiative rates.   Therefore, it is desirable to  measure quantum efficiency, $\gr$ and $\gnr$ independently on individual NV centers  in order to decide if /which  NV centers in nanodiamonds are suitable for use as LDOS probes for integration in nanophotonic devices.  In literature, several reports have appeared that evidence lifetime changes of NV emission  that are induced by placing nanocrystals in photonic environments of varying index and topology~\cite{Schietinger_NL_2009,Kolesov_NPh_2009,Huck_PRL_2011,Ruijgrok_OE_2010,Inam_NJP_2011}.   To our knowledge all these measurements of lifetime changes for nanodiamonds were performed by comparing the \emph{mean rate} from an ensemble of single-center measurements in one system, to measurement on a different ensemble of single-centers in a second system~\cite{Kolesov_NPh_2009,Ruijgrok_OE_2010,Inam_NJP_2011}.  Ruijgrok \etal \cite{Ruijgrok_OE_2010} in particular report changes for a system in which the induced LDOS change is exactly known, and report changes in the \emph{mean} rate consistent with those expected for unit-quantum-efficiency emitters. However, we note that in these ensemble measurements, the change of the \emph{mean} decay rate  was far smaller than the width of the rate distribution.  Therefore, those measurements do not allow to ascertain whether for \emph{any given}  single NV center the rate actually varies with varying LDOS in a manner that is consistent with the expectations for  efficient emitters.

We perform measurements in which we change the LDOS of single NV centers  in two ways.  First, we have immobilized the nanocrystals by evaporating  a thin layer of SiO$_2$, and then introduced  liquids with different refractive indices  to systematically modify the LDOS around the NV centers  (section~\ref{sec:liquidldos}). Second, we have used nanomechanical tuning of LDOS using a piezo-controlled mirror to effectuate a Drexhage experiment~\cite{Drexhage_JL_1970,Leistikow,LodahlDrexhage,AmosBarnes,Kwadrin}  on single nanodiamonds  (section~\ref{sec:mirrorldos}).    The nanomechanical method is preferable since it rapidly allows continuous variation of LDOS.  Due to their  smaller apparent brightness, however, we were unable to apply the technique also to 25\,nm diamond nanocrystals as the much higher  pump energy required to reach appreciable count rates caused thermal breakdown of the micro-mirror. Therefore we applied the liquid tuning method (section \ref{sec:liquidldos}) to the small nanodiamonds, and report on the nanomechanical method for the large nanodiamonds (section \ref{sec:mirrorldos}).

\subsection{Liquid tuning of LDOS on single NV centers in small nanodiamonds\label{sec:liquidldos}}
For the small nanodiamonds  only the liquid tuning method could be applied due to experimental constraints. The idea and first implementation of this method was pioneered by Snoeks \etal~\cite{Snoeks}, and later adopted by e.g.~\cite{deDood,Walters,Ruijgrok_OE_2010}.  Sources are placed in close proximity to a planar interface between a dielectric and a half-space that can be filled with  liquids of different refractive index. The advantage of using liquid tuning of refractive index for sources near an interface is that the LDOS changes near an interface are nearly independent of emission frequency, and are known to be excellently described by the theory explained in full by Urbach and Rikken~\cite{Urbach}. In view of the broad emission spectrum of NV centers in diamond, it is important to apply a broadband LDOS change  when seeking to measure lifetime changes.  A narrowband LDOS variation as obtained with a high-Q microcavity that would for instance be tuned to the zero phonon line,  would not necessarily affect the rate, but rather only the branching ratio between the zero phonon line and the rest of the spectrum.

\begin{figure}[t]
 \includegraphics[width=150mm]{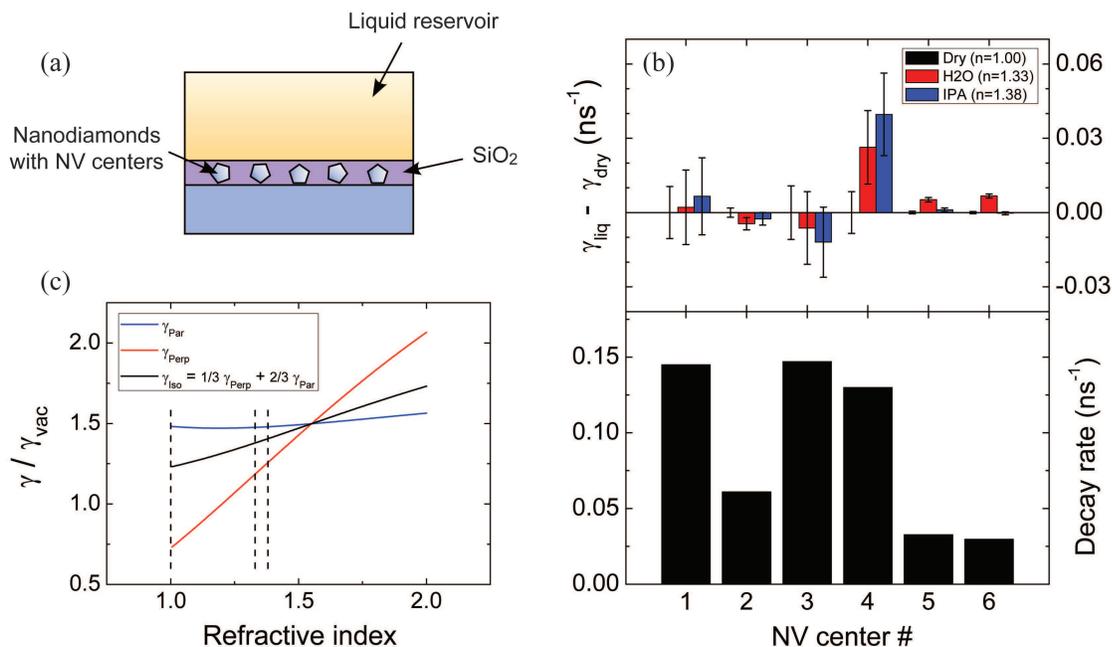}
 \caption{(a) Schematic of the sample used for liquid LDOS tuning experiment. (b) Lower panel: decay rates of six randomly chosen NV centers. Upper panel: decay rate changes of each NV center after addition of liquids. (c) Calculated decay rate variations of a dipole embedded at the center of a three layer system as a function of the refractive index of the top layer for different dipole-moment orientations. Dashed lines indicate the refractive indices of air, water and IPA used in the experiment. \label{Fig:Fig_4}}
\end{figure}

We prepared the sample by evaporating a 60\,nm thick layer of SiO$_{2}$ on one of the 25\,nm nanodiamond samples on which we had identified single NV centers. This step immobilizes the nanodiamonds, and ensures that  liquid application to tune LDOS does not add new chemically induced nonradiative decay channels. Next, we defined a 3\,mm deep liquid reservoir on top of the sample using a ring-shaped enclosure cut from Polydimethylsiloxane (PDMS) bonded to the substrate.  {Figure~\ref{Fig:Fig_4}(a) depicts an schematic of the sample.} In order to investigate the decay dynamics of the NV centers in response to the LDOS variations, we randomly selected six NV centers from the ensemble of figure~\ref{Fig:Fig_3} and measured their lifetimes in three conditions: before adding any liquids where the top half space of nanodiamonds consists of air ($n=1.00$), after adding water with refractive index of $n=1.33$, and after adding isopropyl alcohol (IPA) with refractive index of $n=1.38$ to the liquid bath on top of the nanodiamonds. In the lower panel of figure~\ref{Fig:Fig_4}(b) we plot  the decay rate of each NV center as a black bar measured when the sample is dry. The color bars in the upper panel of figure~\ref{Fig:Fig_4}(b) show the difference between decay rates of each NV center after and before introducing a liquid. We  observe that for most centers, the decay rate either barely varies, or varies non-monotonically with applied index, although generally, one expects to observe an increase in the decay rate of a dipole emitter in a homogenous medium by increasing the refractive index of the environment due to an increase of the LDOS~\cite{Urbach}.

For comparison to the data we calculate the normalized decay rate of a dipole positioned in the middle of a 60\,nm thick SiO$_2$ slab ($n=1.52$) which is sandwiched between a semi-infinite slab of quartz with $n=1.46$ at the bottom and a semi-infinite slab with varying refractive index at the top. Figure~\ref{Fig:Fig_4}(c) shows the calculated decay rate of the dipole as a function of the refractive index of the top layer for different dipole-moment orientations. The calculation shows that as the refractive index increases the decay rate change is expected to be monotonic and increasing and is the largest for a dipole oriented perpendicular to the interface while the change is slight for a dipole-moment parallel to the interface. Comparing the calculated rate changes to the measured decay rates (figure~\ref{Fig:Fig_4}(b)), we find no systematic rate variation for the NV centers within the error bars of our experimental data. The error bar for measurements on individual NV centers is composed of two contributions. First, the error bar contains the uncertainty in the fit to the fluorescence decay trace. Second, the experiment is hampered by the fact that the decay rates of some of the single nanodiamonds was observed to jump slightly between observations, even if no change in environment was applied. These jumps amount to approximately 5\%, and occur on time scales of hours to days. To exclude this effect, we have measured the rate for each liquid and a reference in air on the same day for each NV center to generate the plot of rate changes in  figure~\ref{Fig:Fig_4}(b).

We find that there is no evidence that the rate of \emph{individual} NV centers in 25\,nm diamond nanocrystals actually varies in accordance with the known  applied LDOS changes.  This observation  appears to be at variance with earlier work that evidenced a small change in the mean decay rate of an ensemble of NV centers in a liquid-tuning experiment~\cite{Ruijgrok_OE_2010}. As two caveats, we note that first, in that previous work it was not verified if any such change occurred for each source individually,  and second, that a shortcoming of our experiment  for quantititave interpretation is that   for dipoles located close to an interface (as is the case here), the decay rate enhancement  significantly depends on the unknown dipole-moment orientation of the emitter with respect to the interface. While the fact that we do not know the dipole orientations for each NV center makes it impossible to conclude with certainty that an LDOS effect is absent,  two hypotheses for absence of an LDOS effect could be proposed. First, that the quantum efficiency of 25\,nm diamond nanocrystals is low,  or second, that the very high index diamond shell around the NV centers  intrinsically prevents the NV center from responding to (moderate) changes in the LDOS of the environment it is supposed to probe.  We exclude the latter explanation on basis of a set of  finite element (COMSOL) simulations, in which we calculate the decay rate of a point dipole randomly located in a diamond nanosphere, which in turn is placed at varying distances from a planar interface with materials of various dielectric constants. We find that in all cases, the decay rate of the source simply follows the theory of Urbach and Rikken~\cite{Urbach} for a source without dielectric shell,  multiplied by a pre-factor that is essentially a quasistatic  local field correction factor due to the diamond shell and is independent of the varying LDOS\@.

To conclude, our observation that no NV center responds to LDOS changes together with our earlier correlation plots of brightness versus decay rate (figures~\ref{Fig:Fig_2} and~\ref{Fig:Fig_3}) means that the quantum efficiencies of 25\,nm diamond nanocrystals are  low. A second important conclusion is that single NV centers of 25\,nm size   appear unsuitable to measure LDOS changes  due to the slight jumps in lifetime, unless the entire measurement scheme (including reference measurements) takes less than a few hours.

\subsection{Nanomechanical tuning of LDOS on   large nanodiamonds}\label{sec:mirrorldos}
\begin{figure}[t]
 \includegraphics[width=150mm]{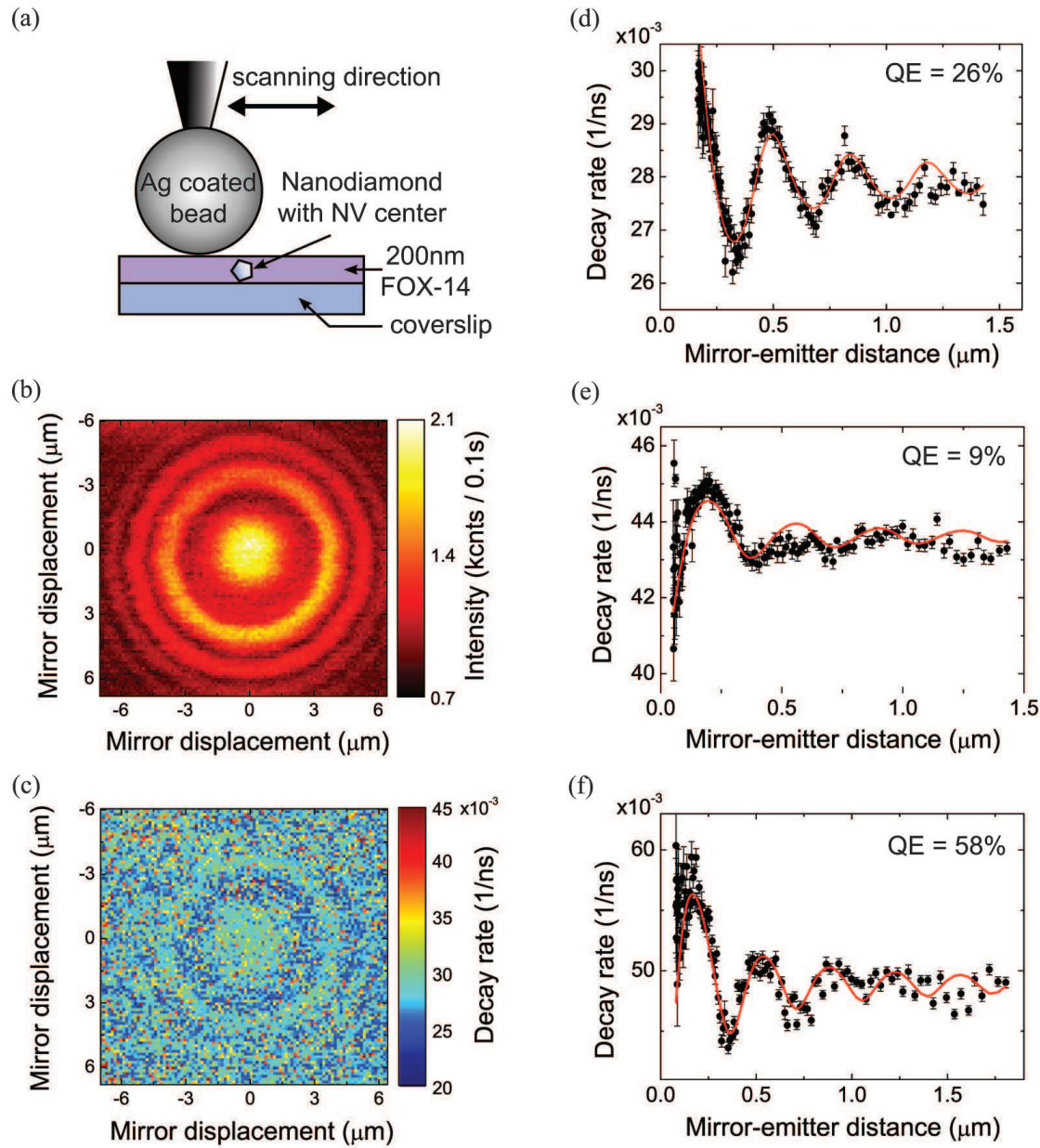}
 \caption{(a) Schematic of the nanomechanical tuning of LDOS for 100\,nm nanodiamonds. (b) Fluorescence intensity map of an NV center (NV center 1 in table~\ref{Table:Table_1}) as a function of mirror lateral displacement (c) Corresponding decay rate map of the NV center as a function of mirror lateral displacement. Decay rates are extracted from a single-exponential fit to the time-trace data for each pixel. (d) Dots: decay rate as a function of the emitter distance to mirror, extracted from the decay rate map shown in (c). Solid line: fitted LDOS on the measured rate oscillations with fit parameters $\gr$, $\gnr$ and a set of dipole moment orientations. The shown QE is the most-likely value. (e) and (f) Decay rate as a function of emitter-mirror distance for two other NV centers (number 4 and 5 in table~\ref{Table:Table_1}, respectively) and the fitted rate from LDOS calculation. \label{Fig:Fig_5}}
\end{figure}

In this section we report on the results of a second method  to assess if single NV centers are suited for LDOS measurements, and to calibrate emission rates. This second method has the advantage that we can apply calibrated LDOS changes rapidly so that jumps in the intrinsic rate constants can be avoided that hamper experiments where macroscopic sample changes are required to modify LDOS, as in a liquid immersion experiment.  The scheme is similar to a measurement procedure reported by  Buchler \etal~\cite{Buchler_PRL_2005}, based on a Drexhage experiment~\cite{Drexhage_JL_1970,Leistikow,LodahlDrexhage,AmosBarnes,Kwadrin}. In a Drexhage experiment, a mirror is used to impose a large LDOS change, and rates are measured as a function of the emitter-mirror separation.   While this method is usually implemented by creating  a set of macroscopic samples where mirrors are coated with spacers of calibrated thicknesses~\cite{Leistikow,LodahlDrexhage,AmosBarnes} or by creating a single sloping wedge between a mirror and an ensemble of emitters~\cite{Kwadrin}, Buchler \etal realized a nanomechanical version that can be applied to a single emitter. Buchler \etal  used a silver-coated curved fiber end as a mirror attached to a piezo-stage to precisely tune the mirror distance to an underlying single emitter on the sample~\cite{Buchler_PRL_2005}. Here, we use a similar method in a shear-force-feedback near-field microscope. However, instead of a vertical mirror displacement, we use a lateral displacement of the curved mirror while staying in shear-force feedback to keep the mirror and sample substrate in near-contact.  As the contact point moves sideways,  the spherical mirror shape ensures that the mirror-to-emitter distance varies (figure~\ref{Fig:Fig_5}(a)). To fabricate the curved mirror, we glued 45\,$\mu$m polystyrene beads (Polysciences, Inc.) to the end of  cleaved optical fibers and coated the beads with about 200\,nm of silver.

We have applied the scanning-mirror LDOS changing technique to 100\,nm nanodiamonds containing single NV centers, prepared as described in section \ref{subsec:smaple_prep}, and  subsequently embedded in a  200\,nm thick layer of planarizing spin-on glass (FOX-14, Dow Corning).  The spin-on glass immobilizes the nanocrystals, so that they are not moved during shear-force scanning.
In order to vary the distance between the mirror and the emitter, we scan the mirror bead laterally on top of an identified NV center. For each position of the mirror bead, we collect the fluorescence emission of the NV center, positioned at the focus of the pump laser, through the confocal microscope setup described in section~\ref{subsec:Exp_setup}. By scanning the mirror bead, we obtain a confocal fluorescence intensity map as shown in figure~\ref{Fig:Fig_5}(b). Here, each pixel represents the relative position of the mirror bead with respect to the NV center with the false color representing the collected fluorescence intensity. We clearly observe  interference rings in the fluorescence intensity map. These rings mainly stem from the fact that the mirror imposes a standing-wave pattern on the 532\,nm pump field, which subsequently results in a modulation of the fluorescence intensity.

For each pixel in the fluorescence intensity map, we stored absolute photon arrival times, as well as laser pulse arrival times, allowing us to extract the decay dynamics. We use a single-exponential fit to obtain the corresponding decay rate of the emitter for each mirror position, as plotted in figure~\ref{Fig:Fig_5}(c). Here, the color scale represents the fitted decay rate corresponding to a defined mirror position. Interestingly,  we  observe a radial modulation of the decay rates varying between about 0.02\,ns$^{-1}$ to 0.04\,ns$^{-1}$, which we  attribute to the varying LDOS in front of the mirror. To quantify the variation as a function of the distance from emitter to mirror, we  extracted the decay rates as a function of  the lateral distance to the mirror central position. To this end, we bin pixels in concentric rings of equal lateral distance to the mirror center, and concatenate the photon-correlation time traces of all pixels in each bin to obtain a single fluorescence decay trace per radial  distance.  Using the spherical form factor of the bead, we convert  lateral position of the mirror to normal distance of the mirror to emitter.
The decay rate fitted to the fluorescence decay for each mirror-emitter separation is depicted in figure~\ref{Fig:Fig_5}(d) (black dots). We observe a distinct oscillation of the rate around 0.028\,ns$^{-1}$, with a 15\% amplitude. To our knowledge, this is the first report of a reversible change in the decay rate of  a single NV center  in a calibrated LDOS experiment.

Due to the large radius of the mirror compared to its distance to emitters, it is reasonable to consider the experimental configuration as a planar glass-air-mirror system in which we vary the air thickness. For this system, the LDOS is exactly known~\cite{Novotnybook,AmosBarnes} for any dipole orientation and position.   We fit the experimental decay rate data using the theoretical LDOS $\rho(z,d,\theta)$ according to $\gamma(z)=\gamma_{nr}+\gamma_r \rho(z,d,\theta)$, where the fit parameters are the nonradiative decay rate $\gamma_{nr}$, the radiative decay rate $\gamma_r$, and the dipole orientation $\theta$ relative to the normal to the plane. Finally a small offset $d$ appears that is due to the unknown distance-of-closest approach in shear-force microscopy, of the order of 15\,nm. Care must be taken that a correlation exists between the dipole orientation $\theta$, and the apparent quantum efficiency $\gamma_r/(\gamma_{nr}+\gamma_r)$ returned by the fit routine, due to the fact that the LDOS for different dipole-moment orientations is similar in qualitative $z$-dependence, but different in oscillation contrast. To overcome this dependency, we examine fits of the rate oscillations   for a set of dipole orientations and   evaluate the  goodness of fits individually.  The goodness of fit is established by measuring in how far the residuals of the fit (point-by-point deviation between data and fit function) are within the data point error bar.

This gives a range of quantum efficiency values for which the fit is consistent with the data. In this paper  we report the range of quantum efficiency values consistent with the data given that  the dipole orientation should be treated as an unknown parameter.  In case the dipole moment orientation of the NV center is known, which could in principle be realized using Fourier microscopy~\cite{JOSALieb},  one can extract a more precise value for the quantum efficiency. For the particular data set shown  in figure~\ref{Fig:Fig_5}(d), we find a most likely quantum efficiency of 26\% and a dipole moment orientation within $20^{\circ}$ along the sample plane. The range of quantum efficiencies  consistent with the data for this NV center is bounded from below by 26\%, and from above by  50\%. To our knowledge this is the first experimental calibration of the quantum efficiency of a single NV center in a diamond nanocrystal.

In order to investigate the typical quantum efficiencies of NV centers, we selected five random NV centers and examined them with the moving mirror experiment as explained above. Each nanodiamond was selected to be a single NV center containing nanocrystal according to the criteria we outlined in section~\ref{Single_NV_charac}, without further  post-selection for inclusion in the Drexhage experiment. Table~\ref{Table:Table_1} summarizes the confidence intervals for the fitted values of the quantum efficiencies, given that we do not know the dipole orientation a priori.  The total decay rate $\gt$, i.e. the sum of $\gr$ and $\gnr$ varies almost over a factor 2, and can be fitted accurately. In Table~\ref{Table:Table_1} we also report radiative and nonradiative decay rates. Values are reported as fitted to the data while fixing the dipole orientation at its most likely value, in which case the data can be fit with small error bars on $\gr$ and $\gnr$. The wide range of quantum efficiencies consistent with the data of course imply that in a completely free fit $\gr$ and $\gnr$ both have a large error bar, though their sum $\gt$ does not. We find that the different NV centers we probed, while having reasonably comparable brightness, in fact have widely different quantum efficiencies.  For instance, NV center 4 (figure~\ref{Fig:Fig_5}(e)) has an efficiency certainly below 14\%, while NV center 5 (figure~\ref{Fig:Fig_5}(f)) has a quantum efficiency certainly above 58\%, in the range 58\% to 90\%. Furthermore, we note that the fits neither result in the conclusion of a fixed nonradiative rate at varying radiative rate, nor conversely in the conclusion that the radiative rate is a constant while the nonradiative rate varies. Instead, both the radiative and the nonradiative decay rates are distributed. Even with the wide error bars on quantum efficiencies, we establish that both distributions  span  at least around a factor two in range. Our measurements hence show that the common tacit  assumption of near-unit quantum efficiency and nearly identical emission characteristics barring those due to depolarization factors in the nanodiamonds should be discarded.
\begin{table}
    \begin{tabular}{ |l || c | c | c | c | c |}
      \hline
      NV center \# & 1 & 2 & 3 & 4 & 5\\
      \hline
      QE range & 26\%\,--\,50\% & 23\%\,--\,45\% & 27\%\,--\,58\% & 9\%\,--\,14\% & 58\%\,--\,90\% \\
      \hline
      $\gt$ (ns$^{-1} \times 10^{-3}$) & 25.7$\pm$0.2 & 29.2$\pm$0.2 & 33.9$\pm$0.4 & 42.2$\pm$0.2 & 39.7$\pm$0.9 \\ \hline   \hline
      $\gr$ (ns$^{-1} \times 10^{-3}$) & 6.7$\pm$0.5 & 6.8$\pm$0.6 & 9$\pm$1 & 4.0$\pm$0.4 & 23$\pm$3 \\
      \hline
      $\gnr$ (ns$^{-1} \times 10^{-3}$) & 19.0$\pm$0.5 & 22.5$\pm$0.6 & 25$\pm$1 & 38.2$\pm$0.5 & 17$\pm$3 \\
      \hline  \hline
      Brightness (kADU/300\,ms) & 102 & 112 & 139 & 144 & 135 \\
      \hline
    \end{tabular}
  \caption{Extracted quantum efficiencies, rates and brightnesses for five randomly chosen NV centers in 100 nm nanodiamonds. Quantum efficiencies are quoted as a range of values, taking into account that the dipole orientation is a free parameter.  In this fit procedure, the proper fit parameters are the quantum efficiency and the total decay rate, which has small error bar.  For completeness we also show the two derived quantities $\gr$ and $\gnr$.   These rates and their error bars are given as fitted when fixing the dipole orientation to its most likely value. Fitted brightnesses were converted to  CCD response units to be comparable with Fig. 3\label{Table:Table_1}}
\end{table}

\section{Conclusion}
In order to assess the suitability of nanodiamonds as photonic LDOS probes and as building blocks for hybrid photonic systems, we have investigated the brightness, decay rate, and quantum efficiency of single NV centers in nanodiamonds of 25 nm, and of 100\,nm median size. For both size ranges a wide distribution in brightnesses and rates is found, consistent with reports by earlier workers.   We conclude that the wide distribution of rates is due to a distribution in radiative rates, nonradiative rates and quantum efficiencies. This conclusion contradicts earlier work, that interpreted  the wide distribution of decay rates as mainly due to a photonic effect that causes a distribution in radiative decay rates via the LDOS. Instead, our measurements show that   even NV centers in large nanocrystals show a wide range of quantum efficiencies when probed in a controlled LDOS experiment.  For NV centers in 100\,nm nanocrystals we find quantum efficiencies distributed between 10\% and 90\%, while for the  smaller NV centers the quantum efficiency  for fluorescence is likely a factor of two smaller on average.

For applications in nanophotonic experiments using single NV centers, a highly problematic property is that quantum efficiency does not correlate with brightness, due to the fact that both the radiative and nonradiative rate are distributed.  In other words, screening nanodiamonds to pick the best one to probe LDOS or for incorporation in a photonic device can not rely on a simple metric such as spectrum or brightness.  Instead, we argue that future work in the hybrid assembly of nanodiamonds in plasmonics and photonic crystals to realize accelerated spontaneous emission decay always requires an experimental protocol in which nanodiamonds are first individually calibrated in terms of quantum efficiency. While we have shown a method for such calibration,  this is a highly tedious procedure that is not easily implementable and requires dedicated  near-field manipulation equipment.

In this work, we have not speculated on the origin of the apparent low quantum efficiency of single NV centers in nanodiamonds. We close with two remarks on the origin of the low apparent quantum efficiency.  Firstly, we have treated the NV center as a quasi two-level system. Our experiments thus address the question what the apparent quantum efficiency is when attempting to use an NV center as a two-level LDOS probe. From the NV center spectrum it is clear that the spectrum is very wide with a large vibrational  broadening spectrum.  Moreover it is well known that the NV$^{-}$ center is not a two level system.  Instead, the NV$^{-}$ has different spin sublevels, and may experience spin-flip intersystem crossing between allowed spin transition manifolds. The rates for these transitions were recently characterized in detail for NV centers in bulk diamond in Ref. \cite{RobledoNJP}.  A further complication is that   the NV$^{-}$ defect may infrequently transition to an uncharged NV center that also luminesces, yet at different rate and efficiency. The many rates involved in these transitions can further vary between NV centers due to, e.g., variations in crystal strain.  A full treatment of the response of NV centers to LDOS changes would hence have to treat the full rate equations in which radiative transition rates are, and intersystem crossing rates are not, affected by LDOS.  An important step beyond our work will be to perform the measurements we have described here but employing the   spin selective techniques reported by Robledo~\cite{RobledoNJP} to establish what the nonradiative and radiative rates are for each transition separately, instead of lumping rates into effective two-level parameters.  While the LDOS changes we have applied using a mirror are broadband LDOS changes that modify  radiative transitions roughly equally across the emission spectrum,  LDOS changes that have strong spectral features could be used advantageously to enhance or suppress the importance of intersystem crossing.

Having established that further work is required to separate the quantum efficiencies reported here into parameters per transition in a more complete level scheme, we turn to possible reasons for the below-unity values of quantum efficiency that we find.   The lower brightness of the smaller NV centers, as well as their lack of response to LDOS changes suggest that the surface, i.e., surface contamination with carbon, or surface defects may play a role in providing quenching sites. Indeed, it  has already been suggested for very small (5\,nm) nanodiamonds that NV centers may suffer quenching due to graphite shells on the diamonds. While the nanodiamonds we used have been employed by several groups in spontaneous emission control experiments in untreated form, exactly as in our experiments,  additional surface treatments such as by prolonger immersion in boiling sulphuric acid,  or cleaning in K\"oningswasser have been proposed by several workers.  Whether or not such treatments actually affect quantum efficiency is as yet unclear, as is whether quenching can be completely suppressed.  We propose that the quantum efficiency calibration method that we demonstrated will be an indispensable tool to evaluate such cleaning methods,  as well as to  screen  other color centers  in diamond for advantageous fluorescence properties.

\ack

We thank S. Schietinger for communications regarding sample preparation methods, M. Frimmer for experimental help and discussions, and C. Osorio for suggestions to improve the manuscript.   This work is part of the research program of the "Foundation for Fundamental Research on Matter (FOM)", which is financially supported by the "The Netherlands Organization for Scientific Research (NWO)". AFK gratefully acknowledges a NWO-Vidi grant for financial support.

\section*{References}

\end{document}